# Edge System Design Using Containers and Unikernels for IoT Applications

Shahidullah Kaiser
*Department of Computer science*
*University of Texas at San Antonio*
San Antonio, TX, USA
shahidullah.kaiser@my.utsa.edu

Ali Şaman Tosun
*Mathematics and Computer Science*
*University of North Carolina at Pembroke*
Pembroke, NC, USA
ali.tosun@uncp.edu

Turgay Korkmaz
*Department of Computer Science*
*The University of Texas at San Antonio*
San Antonio, TX, USA
Turgay.Korkmaz@utsa.edu

*Abstract*—Edge computing is emerging as a key enabler of low-latency, high-efficiency processing for the Internet of Things (IoT) and other real-time applications. To support these demands, containerization has gained traction in edge computing due to its lightweight virtualization and efficient resource management. However, there is currently no established framework to leverage both containers and unikernels on edge devices for optimized IoT deployments. This paper proposes a hybrid edge system design that leverages container and unikernel technologies to optimize resource utilization based on application complexity. Containers are employed for resource-intensive applications, e.g., computer vision, providing faster processing, flexibility, and ease of deployment. In contrast, unikernels are used for lightweight applications, offering enhanced resource performance with minimal overhead. Our system design also incorporates container orchestration to efficiently manage multiple instances across the edge efficiently, ensuring scalability and reliability. We demonstrate our hybrid approach's performance and efficiency advantages through real-world computer vision and data science applications on ARM-powered edge device. Our results demonstrate that this hybrid approach improves resource utilization and reduces latency compared to traditional virtualized solutions. This work provides insights into optimizing edge infrastructures, enabling more efficient and specialized deployment strategies for diverse application workloads.

## I. INTRODUCTION

The rapid growth of the Internet of Things (IoT) has transformed modern computing, generating vast amounts of data that require real-time processing, analysis, and decision-making. Traditional cloud computing solutions, while powerful, are often hindered by high latency and bandwidth limitations, making them less suitable for applications that require immediate processing. Edge computing has emerged as a solution to these challenges by bringing computation closer to data sources, allowing faster response times, and reducing data transmission costs [1]. This shift in computational paradigms is especially relevant for IoT applications that generate diverse data types, such as video streams and sensor readings, which differ significantly in processing demands [2].

In edge computing, leveraging lightweight virtualization technologies has become essential for efficiently managing resource-constrained devices. Containers and unikernels are prominent technologies that provide isolation and streamline deployment on edge devices. Containers like Docker and Podman offer a lightweight alternative to traditional virtual machines, enabling efficient resource sharing and rapid deployment of complex applications [3]. Unikernels, on the other hand, are specialized, single-application kernels that compile only the necessary components required by the application, resulting in minimal overhead and fast boot times. These characteristics make unikernels highly efficient for lightweight and single-purpose applications, such as basic data processing tasks.

To meet the growing demands of modern IoT applications, it is essential to propose efficient and scalable designs that can optimize data processing at the edge. Unlike traditional IoT sensor nodes, edge devices enable faster decision-making, optimize bandwidth usage, and reduce operational costs by processing data closer to its source. However, edge devices require sophisticated architectural designs that can manage diverse workloads in resource-constrained environments to harness these advantages fully. This demand calls for innovative approaches capable of developing advanced edge systems to support the evolving needs of the IoT-Edge ecosystem.

This paper proposes a hybrid edge system architecture that combines containers and unikernels within an ARM-powered edge computing environment to optimize resource allocation and processing efficiency. The contributions of the paper are as follows:

- This paper explores the potential of a hybrid edge computing system by utilizing containers and unikernels to manage various IoT data types efficiently.
- Explores container orchestration frameworks such as Docker Swarm, KubeEdge, K3s, and Nomad to manage instances across multiple edge nodes, further enhancing the scalability of our approach.
- Implementation of a prototype on ARM-based edge devices, evaluating the suitability of this architecture for environments with limited resources.
- Assess the effectiveness of the proposed system by deploying computer vision and data science workloads and evaluating key performance metrics, including processing time, CPU, memory usage, and network latency.

In the following sections, we review relevant background information and related work on edge computing, containers, and unikernels in section II. In section III, we explain our system architecture and experimental setup. We then validate the proposed system and present our findings in section IV. We conclude with insights into future research directions in V.

## II. BACKGROUND AND RELATED WORK

Containers and unikernels are two prominent virtualization technologies explored in edge computing to meet the diverse requirements of resource-constraint environments. Containers offer a lightweight virtualization layer, allowing multiple applications to run isolated within shared OS resources [4]. They are highly portable, scalable, and compatible with orchestration tools, making them popular for managing complex edge workloads. However, containers require an entire OS kernel, which can be resource-intensive for some edge devices with limited computational power and memory. Unikernels, in contrast, are specialized single-purpose VMs that compile only the essential components of an application with a minimal OS footprint. This reduces memory overhead, improves security by minimizing attack surfaces, and provides ultra-fast boot times, making unikernels ideal for specific, lightweight edge applications [5]. However, their lack of compatibility with standard orchestration and their complex development process challenges their standalone use as a platform on edge node.

Recent research has increasingly focused on optimizing containers and unikernels to better suit the demands of edge computing applications [6]–[8]. These studies primarily focus on performance comparison. Jing et al. [9] provide a comparative analysis of Docker and Containerd runtime performance, as well as Kubernetes and Docker Swarm as orchestration tools on IoT devices, assessing their suitability for lightweight, distributed edge tasks. Another notable study by Buyya et al. [10] investigated a lightweight container middleware designed for edge-cloud architectures. They utilize microservices and clustered container environments to overcome orchestration and DevOps challenges traditionally encountered in cloud computing. Morabito's work on performance evaluation across various ARM-based, low-power devices further underscores the resource-constrained nature of edge environments [11]. His study tested Docker on Raspberry Pi models 2 and 3 and multiple Odroid devices. He measures key metrics such as CPU usage, power consumption, memory usage, and network and disk performance, which provide insights into container suitability for low-power edge deployments.

Container orchestration is essential for efficiently managing distributed applications across various edge nodes when multiple containers share resources. It automates container deployment, scaling, and monitoring, ensuring optimal resource use and reliability. Seong et al. [12] evaluated KubeEdge's resource distribution and latency performance, proposing a local scheduling scheme that showed improvements over standard load-balancing algorithms. Kjorveziroski et al. [13] compared Kubernetes, K3s, and MicroK8s in resource-constrained settings with 14 benchmarks, finding that K3s and MicroK8s generally offered better performance, although Kubernetes excelled under sustained loads. Böhm et al. [14] similarly noted that K3s achieved better resource utilization than full Kubernetes. Bahy et al. [15] implemented K3s and Nomad on ARM and x86 platforms, analyzing CPU, memory, and storage performance for edge scenarios. Other studies have conducted qualitative and quantitative evaluations of Kubernetes networking solutions in resource-limited environments, highlighting the advantages of lightweight Kubernetes distributions for edge computing [16], [17].

There is growing interest in hybrid approaches combining the benefits of containers and unikernels to optimize performance and resource efficiency in edge computing. This approach leverages containers for complex, multi-purpose applications and unikernels for lightweight, single-purpose tasks.

## III. EDGE SYSTEM DESIGN

In designing the upper-level system architecture, we selected and developed a solution that best aligns with our system's goals and requirements. This section provides an overview of the architecture specifications and the technologies employed, covering both hardware and software perspectives. Our proposed system's overview is shown in figure 1.

### A. Edge System Requirement

Our edge system design is structured around two core principles: Resource-awareness and Application-awareness. These principles guide how tasks are allocated and processed across containers and unikernels.

**Resource-aware:** The system actively monitors available resources on each edge device, such as memory and storage. This real-time awareness allows the system to dynamically distribute computational tasks and manage data storage based on resource availability. By prioritizing tasks that can be accommodated within the current resource constraints, the system minimizes the risk of overloading edge nodes, ensuring stable performance even under heavy workloads.

**Application-aware:** The edge controller categorizes incoming tasks based on application type, determining whether they are best suited for containerized environments or unikernels. For instance, computationally intensive tasks, like those involved in computer vision or complex data analysis, are directed to containerized environments. Lightweight applications, such as sensor data processing, are assigned to unikernels, taking advantage of their efficiency in resource-constrained settings. This selective deployment enhances the system's adaptability, saves resources, and allows it to optimize performance according to the specific requirements of each application.

By integrating resource and application awareness, this edge system design supports a more intelligent, efficient, and adaptable framework for IoT-Edge ecosystems, capable of handling diverse workloads with optimized resource utilization.

### B. System Design

Figure 2 shows a schematic representation of our edge system. The system design incorporates a configuration manager

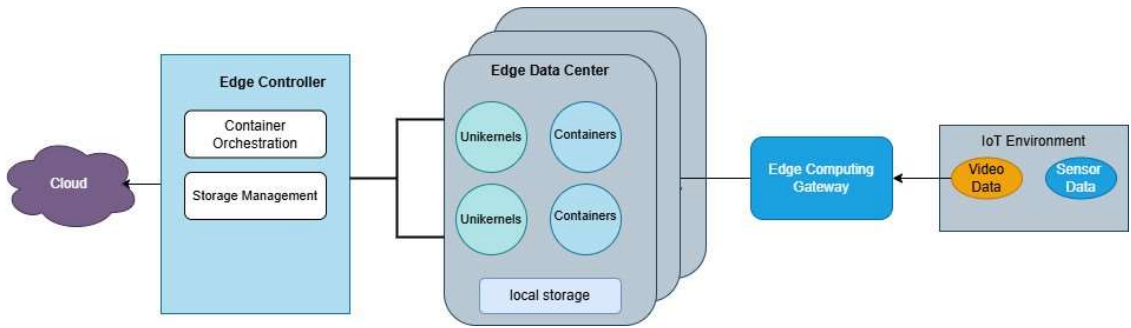

Fig. 1: Overview of our proposed approach

that dynamically allocates resources based on data type and application needs.

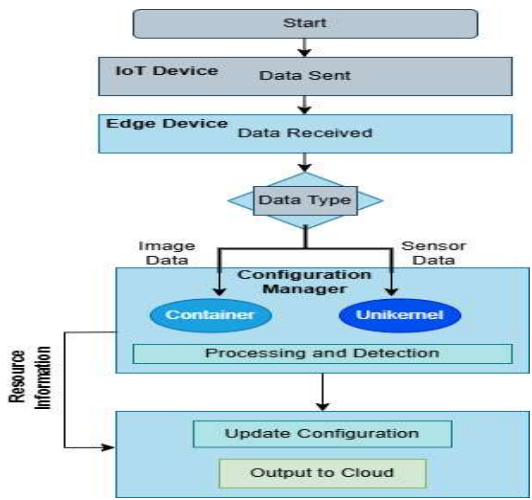

Fig. 2: Schematic representation of our Edge System Architecture

It also orchestrates containers for running multiple instances on the manager and worker nodes. The configuration manager optimizes resource allocation dynamically, ensuring the system can handle multiple simultaneous tasks with minimal performance degradation.

### C. Container and Unikernel

We implement Docker [18], Podman [19], and Singularity [20] as container runtimes on ARM-based edge devices to evaluate our proposed system design. We focus on their suitability for handling computer vision tasks and data science workloads often encountered in IoT-Edge scenarios. Our experiments included applications such as Haar Cascades, HOG, and CNN with the YOLO algorithm. Additionally, to test lightweight tasks on unikernels, we process Fitbit fitness tracker data using Unikraft [21], OSv [22], and Nanos [23], assessing their ability to handle simpler edge workloads in a highly efficient manner.

### D. System Hardware

The hardware setup for our edge computing system leverages Raspberry Pi4. The edge cluster consists of five Raspberry Pi4. Each Raspberry Pi4 is connected to WiFi connections. We assign one node as manager and the other four as worker nodes. Table I gives an overview of the specification of our experimental setup for each node.

| Description | Specification |
| --- | --- |
| Architecture | ARM 64 |
| SoC | Broadcom BCM2711 |
| Processor | ARM 1.5GHz CPU |
| Total Cores | Quad core Cortex A-72 |
| Total RAM | 4GB |
| OS | Linux |
| Power Consumption | 500 mA |
| Micro-SD card | Available |
| Wireless | 2.4 GHz and 5GHz |
| Bluetooth | 5.0 |

TABLE I: Specification of Experimental Setup

For our experiment, we install Ubuntu 22.04, a 64-bit operating system. An ASUS RT-N56U Dual Band Wireless N600 Gigabit router facilitates network setup. This router features 802.11n wireless capabilities and LAN ports with up to 1000 Mbps, providing reliable connectivity for testing high-bandwidth and latency-sensitive applications.

### E. Load Balancer

Container orchestrators enable streamlined management by automating deployment and scaling across multiple devices and networks. This is particularly useful in network failures or bandwidth limitations scenarios, as containers can be quickly redeployed to alternate devices, ensuring uninterrupted service. Additionally, orchestration allows dynamic load balancing. During periods of high load, additional containers can be deployed across multiple devices to balance the workload efficiently. Conversely, scaling down the number of active containers in low-load situations can help conserve energy and reduce operational costs.

In our system, we selected different orchestrators as load balancers based on their unique strengths for edge environments. Docker Swarm [24] offers simplicity and ease of use, making it ideal for straightforward deployments that require basic load-balancing capabilities. KubeEdge [25], explicitly built for edge and IoT use cases, brings robust Kubernetes compatibility, allowing consistent orchestration across cloud and edge environments. K3s [26], a lightweight version of Kubernetes, is well-suited to resource-constrained edge devices, providing essential orchestration functions without the overhead of full Kubernetes. Lastly, we experiment with Nomad [27] orchestrator, a highly flexible orchestration tool that supports many use cases and offers extensive configuration options.

## IV. System Design Validation

We evaluate our system architecture within a real-world IoT-Edge environment. The evaluation begins with IoT devices transmitting data to edge nodes where containers and unikernels are deployed. The configuration manager identifies the data type and allocates tasks accordingly. If the incoming data is an image, it is processed by a container. Upon receiving the image, the container performs detection, saves the output image by category (e.g., Face, Vehicle, Body, or Object) in JPG format, and logs resource usage and processing time. We obtain the images in our experiments from two video clips [28], [29]. For stream data, the task is directed to a unikernel. To simulate real-time processing, we use Fitbit fitness tracker data [30], specifically the Daily Activity dataset, which includes attributes like *ActivityDate, TotalSteps, TotalDistance,* and *Calories*. Each unikernel records key metrics such as processing time, CPU usage, and RAM utilization.

### A. Container Deployment

Figure 3 presents the results of our container-based system validation using computer vision applications. Docker shows the lowest CPU consumption for Car detection (26%) and delivered near-native performance (25.03%). However, for more computationally demanding applications like object detection, which relies on deep neural networks, singularity outperforms other containers by handling higher CPU demands more efficiently. This finding highlights the need for optimized algorithms and container solutions in edge environments to enhance CPU efficiency.

RAM usage was evaluated across different containers for each application. For Face and Car detection, the containers consumed a similar range of RAM, between 90-96 MB, while the native system used around 79 MB. The Body detection application required slightly less memory, with an average usage of 80-84 MB. However, RAM consumption increased significantly for the Object detection application, primarily due to the resource-intensive deep neural network algorithms employed. These results underscore the impact of application complexity on resource usage and the importance of selecting appropriate container technologies for optimized performance in edge computing.

### B. Unikernel Deployment

Our experiments examined resource utilization on an edge node running unikernel-based deployments. Figure 4 illustrates the CPU and RAM usage for Unikraft, Nanos, and OSv when processing stream data. Among the unikernels, Unikraft demonstrated the lowest CPU consumption, with values tightly clustered between 0.17% and 0.20%. OSv exhibited a slightly higher CPU usage, ranging from 0.19% to 0.26%, while Nanos showed moderate CPU consumption between 0.19% and 0.24%. In terms of memory efficiency, Unikraft again proved the most lightweight, utilizing approximately 45-48 MB of RAM. Nanos followed closely, maintaining efficient memory usage with an average of around 50 MB. OSv, however, recorded the highest memory usage among the three, averaging around 55 MB. These findings indicate that Unikraft is particularly effective for resource-constrained edge environments with critical CPU and memory usage. However, Unikernel is not yet ready for any image processing task where computer vision applications need to run.

We run the same data science application on containers to assess how a hybrid edge system can conserve resources such as CPU and memory. Figure 5 illustrates that the Docker container consumes an average of .29% CPU and 71 MB memory, compared to .17% CPU and 45 MB RAM by Unikraft. Thus, using a unikernel saves approximately 36.62% of memory compared to a container. These findings underscore the resource efficiency of using unikernels for lightweight tasks. By selectively deploying unikernels for less resource-demanding applications rather than running all tasks in containers, we can effectively reduce resource usage on the edge node, enhancing the system's overall efficiency. However, there is a trade-off between resource usage and execution time, which we will examine in the following section.

We further validate our edge system design using metrics such as processing time, which is crucial in edge computing, impacting real-time responsiveness, low latency, and bandwidth efficiency. For computer vision applications, the average processing or execution time defines how much time containers need to identify an image correctly. In the case of stream data, it is the time to receive the data and perform the given task: calculate the average steps per user and find the maximum average steps. Figure 6a demonstrates the processing time for the containerized computer vision applications. The average processing time for Car, Face, and Body detection is .12, .2, and .4 seconds, respectively. For object detection, Docker spent the most time among containers for processing, with an average of 1.3 seconds; however, Singularity and Podman performed better in this case.

Figure 6b shows that Unikraft takes around 2 to 2.1 msec. Nanos and OSv exhibit higher processing times, with OSv being the slowest, crossing 2.5 msec for most trials. On the contrary, Figure 6c shows that the Docker container takes 1.7 msec and the Singularity container takes only 1.503 msec, the lowest among all the containers. This experiment shows the trade-off between Container and Unikernel on Data science workload. Container consumes more resources; however, they have faster processing time. On the contrary, Unikernel consumes lower resources, but processing time is comparatively slower.

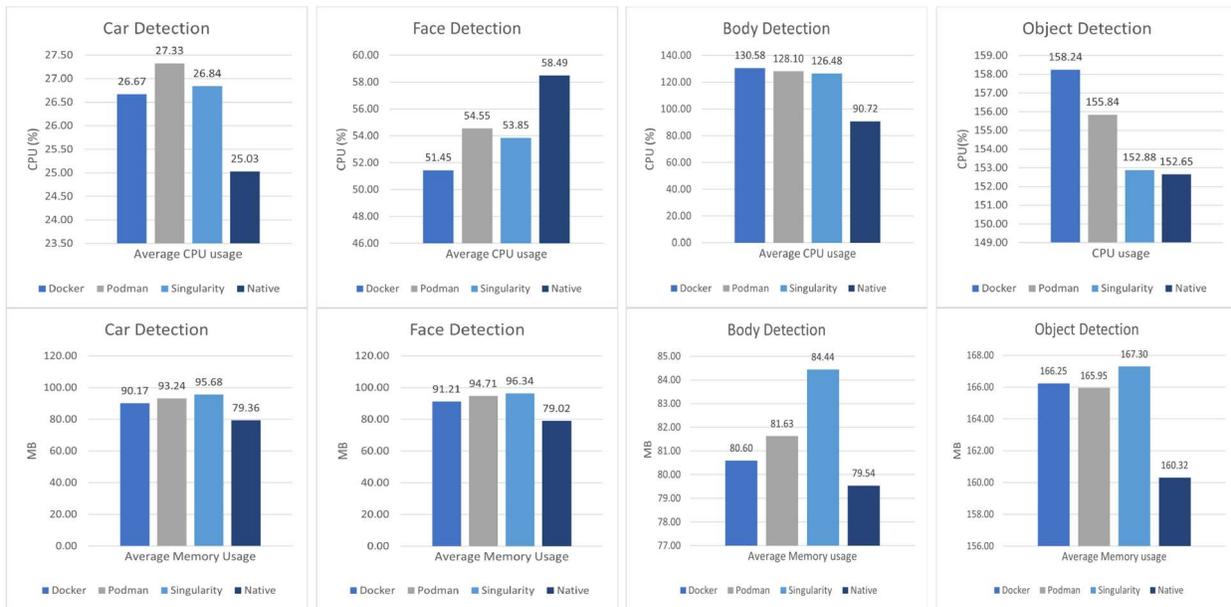

Fig. 3: Average Resource usage of selected container for Computer Vision Application

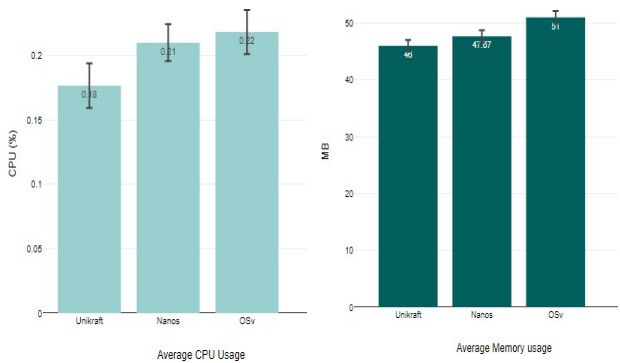

Fig. 4: Average Resource usage of selected Unikernel for Data Science Application

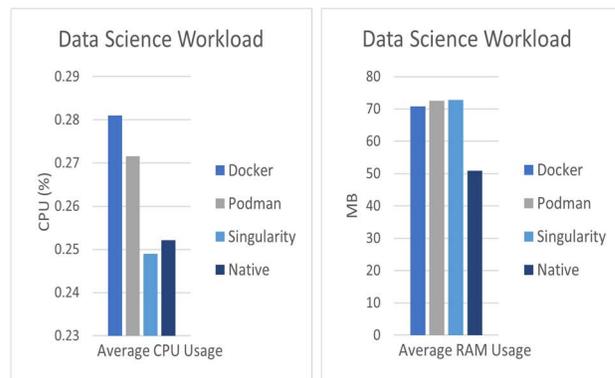

Fig. 5: Average Resource usage of selected Container for Data Science Application

*C. Container Orchestration Deployment*

We further validate our system design by deploying container orchestration tools on edge clusters. Sixteen instances of a computer vision application were deployed across four worker nodes, with the master node located on a separate node. As shown in Figure 7, the average resource consumption was monitored across the cluster. Utilizing orchestration proves advantageous for load balancing, a critical factor in large-scale edge deployments. When a node becomes overloaded with tasks, the manager node dynamically redistributes workloads to other nodes, ensuring balanced resource usage and maintaining system efficiency under high-demand conditions.

## V. CONCLUSION

This study proposes a hybrid edge computing architecture leveraging container and unikernel technologies to enhance workload management for diverse applications. Our architecture demonstrates significant efficiency gains in real-world IoT-edge environments by deploying containers for high-complexity, resource-intensive tasks, such as image recognition, and unikernels for lightweight data processing. The evaluation confirmed that this hybrid model supports optimized resource allocation and responsiveness, addressing critical needs in edge computing scenarios.

Future work will automate the control manager to dynamically assign tasks to containers or unikernels based on application classification, enhancing the system's scalability and adaptability. Moreover, we plan to build and experiment on ARM-based heterogeneous clusters using Raspberry Pi, Rock Pi, and Jetson Nano.

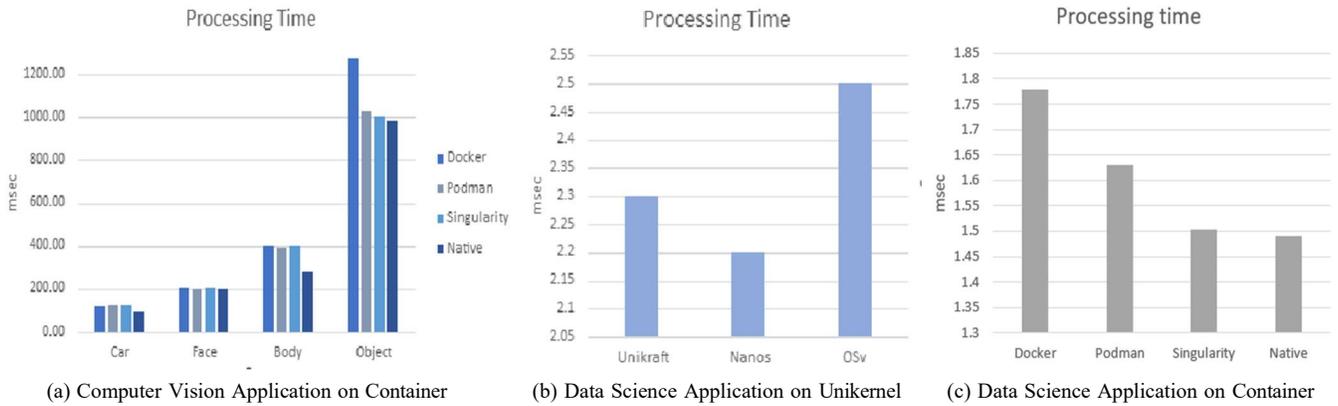

(a) Computer Vision Application on Container   (b) Data Science Application on Unikernel   (c) Data Science Application on Container

Fig. 6: Data Processing Time for Container and Unikernel (lower is better)

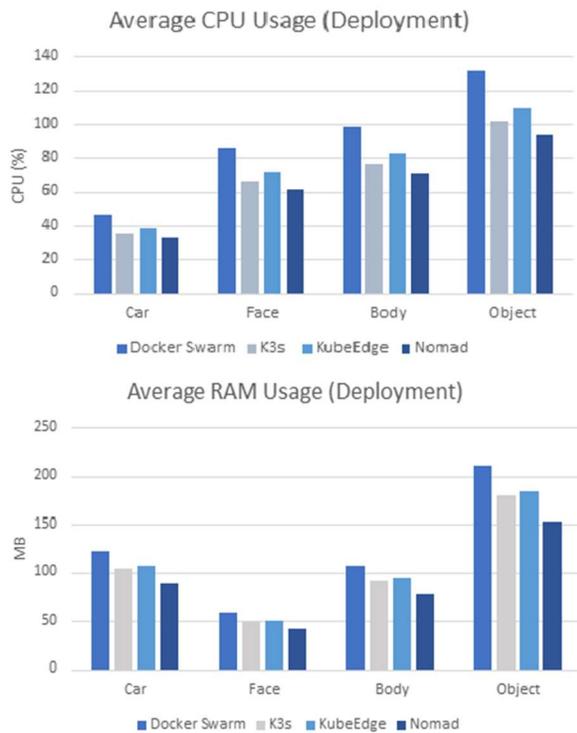

Fig. 7: Resource usage on Container Orchestration Tool